\documentclass{aa}

\newcommand{\magpt}[2]{\mbox{$\rm #1\hspace{-0.25em}\stackrel{m}{.}
      \hspace{-1.0mm}#2$}}                             
\newcommand{\magn}[1]{\mbox{$\rm #1\hspace{-0.05em}^m$}}
\def\bsec{\hbox{$.\!\!{\arcsec}$}}

\newcommand\RA[4]{#1$^{\rm h}$#2$^{\rm m}$#3$\stackrel{\rm s}{.}$#4}

\newcommand\DEC[4]{#1$^{\circ}$#2\arcmin#3\bsec#4}
\newcommand\teff{$ {\rm T_{eff}}$}

\newcommand\logg{$\log {\rm g}$}
\newcommand\loghe{${\rm \log{\frac{n_{He}}{n_{H}}}}$}

\def\gtrsim{\mathrel{\hbox{\rlap{\hbox{\lower4pt\hbox{$\sim$}}}\hbox{$>$}}}}
\def\lesssim{\mathrel{\hbox{\rlap{\hbox{\lower4pt\hbox{$\sim$}}}\hbox{$<$}}}}

\newcommand{\Msolar}{\mbox{\,$\rm M_{\odot}$}}        
\begin{document}
\title{Spectroscopic Analyses of the ``Blue Hook'' Stars in $\omega$
Centauri: A Test of the Late Hot Flasher Scenario}
 \author{S. Moehler\inst{1,2}
\thanks{Based on observations collected at the
European Southern Observatory, Chile (ESO proposal 66.D-0199(A))}
 \and A.V.~Sweigart\inst{3}
 \and W.B.~Landsman\inst{4}
\and S. Dreizler\inst{5}}
\offprints{S. Moehler (Kiel)}
\institute{Dr. Remeis-Sternwarte, Astronomisches Institut der Universit\"at
Erlangen-N\"urnberg, Sternwartstr. 7, 96049 Bamberg, Germany
\and Institut f\"ur Theoretische Physik und Astrophysik der Universit\"at Kiel,
 Abteilung Astrophysik, 24098 Kiel, Germany
(e-mail: moehler@astrophysik.uni-kiel.de)
 \and NASA\,Goddard Space Flight Center, Code 681, Greenbelt,
 MD 20771, USA (e-mail: sweigart@bach.gsfc.nasa.gov),
 \and SSAI, NASA\,Goddard Space Flight Center, Code 681, Greenbelt,
 MD 20771, USA (e-mail: landsman@mpb.gsfc.nasa.gov)
\and
Astronomisches Institut der Universit\"at T\"ubingen,
Sand 1, D-72076 T\"ubingen, Germany
(e-mail: dreizler@astro.uni-tuebingen.de)}
\titlerunning{``Blue Hook'' Stars in $\omega$ Centauri}
\authorrunning{Moehler, Sweigart, Landsman, Dreizler}
\date{Received 3 June 2002 / Accepted 28 August 2002}
\abstract{ $\omega$ Cen contains the largest population of very hot
horizontal branch (HB) stars known in a globular cluster. Recent UV
observations (Whitney et al.\ \cite{whro98}; D'Cruz et al.\
\cite{dcoc00}) show a significant population of hot stars below the
zero-age horizontal branch (``blue hook'' stars), which cannot be
explained by canonical stellar evolution.  Stars which suffer
unusually large mass loss on the red giant branch and thus experience
the helium core flash while descending the white dwarf cooling curve
could populate this region.  Theory predicts that these ``late hot flashers'' 
should show
higher temperatures than the hottest canonical HB stars and should have
helium- and carbon-rich atmospheres.  We obtained and analysed medium
resolution spectra of a sample of blue hook stars to derive their
atmospheric parameters. The blue hook stars are indeed both
hotter (\teff $\ge$35,000~K) and more helium-rich than classical
extreme HB stars. In addition we find indications for a large
enhancement of the carbon abundance relative to the cluster abundance.
}
\maketitle

\section{Introduction}
\begin{figure}[h]
\vspace*{9.cm}
\includegraphics{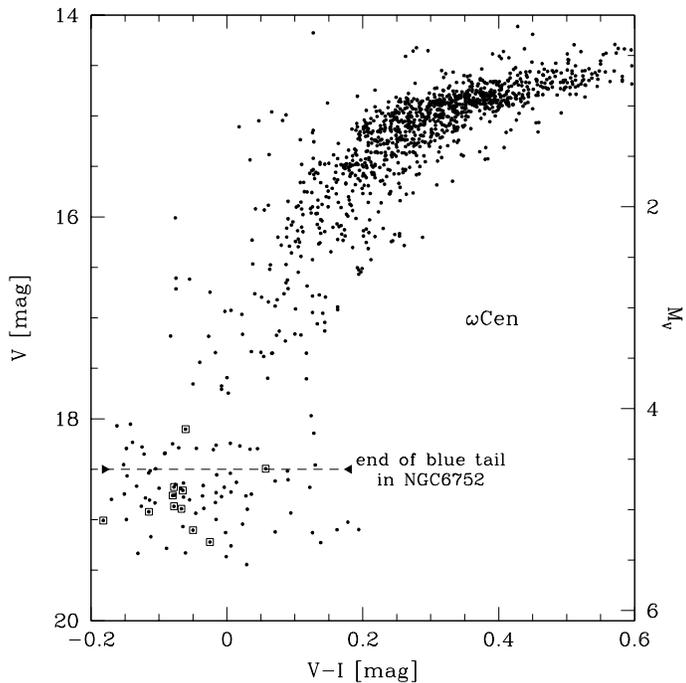}
\caption{
Colour-magnitude diagram of the blue tail of $\omega$ Cen (Kaluzny et al.
\cite{kaku97}) with the spectroscopic targets marked. To establish the
absolute magnitude scale on the right side apparent distance moduli 
$(m-M)_V$ of \magpt{13}{9} and \magpt{13}{3} were used for $\omega$ Cen
and NGC~6752, respectively.
\label{ocen_kalu}}
\end{figure}

Horizontal-branch (HB) stars consist of a helium-burning core of about
0.5~M$_\odot$ surrounded by a hydrogen-burning shell and a
hydrogen-rich envelope of varying mass.  The temperature of an HB star
(at a given metallicity) is determined by the mass of its hydrogen
envelope, with the envelopes of the cooler HB stars being more massive.
The increase in the bolometric
correction with increasing temperature turns the blue HB into a vertical
blue tail in optical colour-magnitude diagrams (CMDs,
cf. Fig.~\ref{ocen_kalu}) with the faintest blue tail stars being the
hottest and least massive.
The hottest HB stars (so-called extreme HB or EHB stars) with $T_{\rm
eff}$ $>$ 20,000 K have so little envelope mass that they are unable
to sustain hydrogen-shell burning.  Nearly all of their surface
luminosity comes from helium burning in the core.  Such EHB stars can
be identified with the subdwarf B (sdB) stars in the field of the Milky Way
and are believed to be mainly responsible for the UV excess observed
in the spectra of elliptical galaxies. The globular cluster $\omega$~Cen
possesses an especially long blue tail containing the largest known
population of EHB stars in a globular cluster.

Observations of $\omega$~Cen in the far-UV (Whitney et al.
\cite{whro98}; D'Cruz et al. \cite{dcoc00}) revealed a
puzzling feature: the very hot end of the HB shows a surprisingly
large spread in UV brightness, including a substantial population
of subluminous stars lying up to \magpt{0}{7} below the zero-age HB
(ZAHB).  Such subluminous EHB stars are so far known to exist only in one
other globular cluster (NGC~2808; Brown et al. \cite{brsw01};
Sweigart et al. \cite{swbr02}).  While stars brighter than the
ZAHB can be produced by evolution away from the ZAHB, the stars
fainter than the ZAHB cannot be explained by canonical HB
evolution.  Within the framework of canonical HB theory there is
no way to populate this region of the UV CMD without requiring an
implausibly large decrease in the helium-core mass.  The subluminous
EHB stars appear to form a hook-like feature in the UV CMD and
are therefore called ``blue hook'' stars.  In optical CMDs these
stars show up at the very faint end of the blue tail (cf. 
Fig.~\ref{ocen_kalu}), in
agreement with the high temperatures suggested by their UV
photometry.

The blue hook stars in $\omega$~Cen populate a range in
absolute visual magnitude that extends beyond the faint limit of
the long blue tail in NGC~6752, which has been studied extensively by
Moehler et al. (\cite{mosw00}).  That in itself would not be a problem,
but the spectroscopic analyses of Moehler et al. (\cite{mosw00}) show that
the blue tail stars in NGC~6752 already populate the EHB to the
hot end predicted by canonical HB models.  Thus canonical theory
fails to explain both the faint UV luminosities and expected high
temperatures of the blue hook stars.  One might suspect that
hotter EHB stars could be produced by simply reducing the
envelope mass even further.  However, Brown et al. (\cite{brsw01}) have
demonstrated that there is a lower limit to the envelope mass of
canonical EHB stars.  Increasing the mass loss along the red-giant
branch (RGB) will not reduce the envelope mass below this
limit but instead will cause a star to die as a helium white
dwarf without ever igniting helium in its core.  Thus the 
blue hook stars may represent a new evolutionary channel for
populating the very hot end of the HB.

One possibility is that the blue hook stars have undergone a
delayed helium-core flash.  Castellani \& Castellani (\cite{caca93}) were
the first to suggest that - for very high mass loss on the
RGB - the helium flash can occur at high effective temperatures after a
star has left the RGB (the so-called ``hot flashers'').  Indeed,
D'Cruz et al. (\cite{dcdo96}, \cite{dcoc00}) proposed that the blue hook stars
could be the progeny of such hot flashers, but unfortunately
the D'Cruz et al. models were, at most, only $\approx$\magpt{0}{1}
fainter than the canonical ZAHB, much less than required by the
observations.  More recently, Brown et al. (\cite{brsw01}) have explored
the evolution of the hot flashers through the helium flash to
the EHB in more detail, especially in regard to the timing of the
flash.  Their models show that under some circumstances the
helium flash will induce substantial mixing between the hydrogen
envelope and helium core, leading to helium-rich EHB stars that
are much hotter than canonical ones.  Brown et al. (\cite{brsw01}) suggest
that this ``flash mixing'' may be the key for understanding the
evolutionary status of the blue hook stars.  Such mixing may also
be responsible for producing the helium-rich, high gravity field
sdO stars (Lemke et al. \cite{lehe97}), whose origin is
otherwise obscure.

The purpose of this paper is to present a spectroscopic analysis
of a sample of blue hook stars in $\omega$~Cen in order to
test the predictions of the flash-mixing scenario.  Following a
brief description of this scenario in Sect.~2, we discuss our
observational data and then derive the parameters of the blue hook
stars (temperatures, gravities and helium abundances) in
Sects. 3 and 4, respectively.  In Sect.~5 we compare our
results with the predictions of the flash-mixing scenario. 

\section{Flash-Mixing Scenario}

In discussing the flash-mixing scenario it is essential to
distinguish between ``early'' and ``late'' hot flashers, since
their evolution through the helium flash differs in fundamental
respects.  The early hot flashers are stars which ignite helium
at some point between the tip of the RGB and the top of the
helium white dwarf cooling curve, i.e., before the ``knee''
visible in Fig. 2, and therefore at a time when the hydrogen-burning
shell is a strong energy source in the star (see long-dashed line
in Fig. 2).  As shown by Iben (\cite{iben76}), a strong hydrogen-burning
shell poses a formidable entropy barrier that effectively
prevents the convection zone produced by the helium flash from
penetrating into the hydrogen envelope.  Thus an early hot
flasher will settle onto the EHB without any mixing between its
helium core and hydrogen envelope and consequently without any
change in its envelope mass or composition.  In other words such
a star will follow a canonical (i.e., unmixed) evolutionary path
to the EHB.  The models of Brown et al. (\cite{brsw01}), which followed
the evolution of the early hot flashers through the helium flash,
showed that these stars reach a maximum temperature of
 $\approx$31,500~K on the EHB.  A similar maximum temperature
is evident in the models of D'Cruz et al. (\cite{dcdo96}, their Fig. 2),
who assumed that the helium flash had no effect on the envelopes
of the EHB stars.  Since the core masses of the early hot
flashers are at most only $\approx$0.001~M$_\odot$ smaller
than the core masses of the EHB stars which ignite helium on
the RGB, their luminosities will be nearly indistinguishable from
the luminosities of the canonical EHB stars.  While such
early hot flashers would populate the clump seen in the UV CMD of
$\omega$~Cen at $m_{160} - V <$ \magpt{-3}{0} and \magpt{14}{8}
$< m_{160} <$ \magpt{15}{3} (D'Cruz et al. \cite{dcoc00}), they are
too bright to explain the blue hook stars.

\begin{figure}[h]
\vspace*{6cm}
\includegraphics{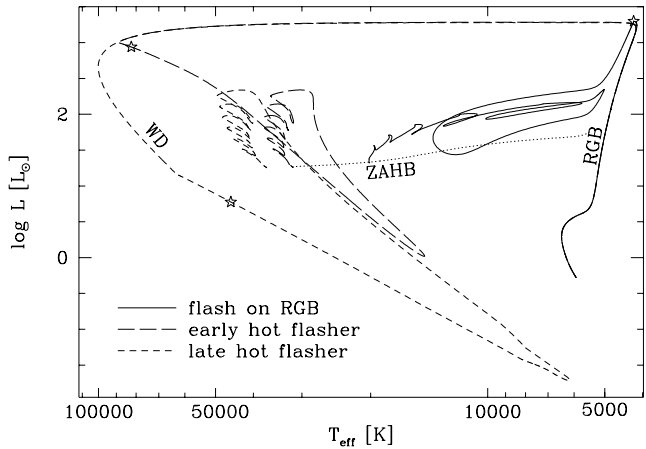}
\caption{ Evolutionary tracks through the helium flash for a star
which ignites helium on the RGB (solid line), an early hot flasher
(long-dashed line), and a late hot flasher (short-dashed line). The peak
of the helium flash along each track is indicated by an asterisk.  The 
dotted line marks the canonical zero-age HB (ZAHB).  The early hot flasher
in this figure produces an EHB star near the hot end of the canonical
HB.  Note the temperature gap between this early hot flasher
and the late hot flasher.  All tracks are taken from
Sweigart (1997).\label{flasher}}
\end{figure}

The evolution of a hot flasher is dramatically different if the
helium flash is delayed until the star is descending the
white dwarf cooling curve (late hot flasher; short-dashed line in
Fig. 2).  The hydrogen-burning shell in a late hot flasher is
substantially weaker than in an early hot flasher, and
consequently it presents a much lower entropy barrier between the
core and the envelope.  As a result, the flash convection is then
able to penetrate through the hydrogen shell into the envelope,
thereby mixing hydrogen from the envelope into the core where it is
rapidly burned (Sweigart \cite{swei97}).  At the same time helium and carbon
from the core are transported outward into the envelope.  This flash
mixing is similar to the mixing that occurs during a very late
helium-shell flash according to the ``born-again'' scenario for
producing H-deficient post-asymptotic giant branch (post-AGB)
stars (e.g., Iben et al. \cite{ibka83}; Iben \cite{iben84},
\cite{iben95}; Renzini \cite{renz90}; Herwig \cite{herw01}).  Similar
mixing has also been found during the helium flash in Population
III stars, where the flash occurs at a much lower luminosity than
in globular cluster stars (Hollowell et al. \cite{hoib90}, Fujimoto et
al. \cite{fuib90}, \cite{fuik00}, Schlattl et al. \cite{scca01}).

The calculations of Brown et al. (\cite{brsw01}) for the late hot flashers
were stopped at the onset of flash mixing due to the numerical
difficulty of following the time-dependent convective mixing of
the envelope hydrogen and the simultaneous nucleosynthesis within
the flash convection zone.  Based on the earlier models of
Sweigart (\cite{swei97}), Brown et al. (\cite{brsw01}) predicted that the flash
convection would capture essentially all of the hydrogen
envelope, thus resulting in a final envelope composition that is
highly enriched in helium and triple-alpha carbon.  This
prediction has recently been confirmed by the detailed
calculations of Schlattl \& Weiss (2002, priv. comm.),
who evolved two late hot flashers through the helium flash to the EHB
using a diffusion algorithm for coupling the nucleosynthesis
to the convective mixing. Flash
mixing in these late hot flashers reduced the envelope hydrogen
abundance X to $\approx$10$^{-4}$ while increasing the
envelope carbon abundance to $\approx$0.03 by mass.  Thus only
a small residual amount of the envelope hydrogen survives the
flash-mixing phase.

Flash mixing introduces a dichotomy in the properties of the EHB
stars that can be observationally tested.  The models of Brown et
al. (\cite{brsw01}) show that the late hot flashers will lie at effective
temperatures of about 37,000~K on the EHB, considerably hotter
than the early hot flashers.  Moreover, the transition between the
early and late hot flashers is exceedingly sharp, corresponding
to a difference in mass loss along the RGB of only
10$^{-4}$\Msolar.  Thus one would expect the large
temperature difference between the early and late hot flashers to
produce a gap in the observed stellar distribution towards
the hot end of the blue tail, as is, in fact, seen in optical
CMDs of NGC~2808 (Walker \cite{walk99}, Bedin et al. \cite{bepi00}). 
 We also note that the change in the surface
composition of the late hot flashers from hydrogen-rich to
helium/carbon-rich reduces the atmospheric opacity below
912~\AA~so that more of the flux is radiated in the far-UV and
less at longer wavelengths.  This effect together with the larger
bolometric corrections of the late hot flashers resulting from
their higher effective temperatures lowers their UV luminosity by
$\approx$\magpt{0}{7}, just as observed in the blue hook stars
(Brown et al. \cite{brsw01}).  This further strengthens the
argument that the blue hook stars are indeed the progeny of the
late hot flashers.

In the next section we present our observational data for testing
the above predictions of the flash-mixing scenario by determining the
temperatures and surface compositions of the blue hook stars.

\section{Observations and Data Reduction}

Due to the predicted change in surface composition the obvious way to
verify the existence of late hot flashers is by spectroscopic
observations of the blue hook stars in $\omega$ Cen.  We obtained
medium-resolution spectra ($R\approx$700) of 12 blue hook candidates
with 18.5 $< V <$ 19.2 at the NTT with EMMI on February 22--25, 2001.
The candidate blue hook stars in the WFPC2 photometry of D'Cruz et
al.\ (2000) were generally too crowded for ground-based spectroscopy,
although we were able to observe one star on the WF3 chip of the
least-crowded pointing (WF3-1).  Our remaining 11 targets come from
either fields BC or C of Kaluzny et al. (\cite{kaku96}) or field D of
Kaluzny et al.  (\cite{kaku97}).  For all the Kaluzny stars we derived
1520 \AA\ photometry from an image of $\omega$ Cen obtained with the
Ultraviolet Imaging Telescope (UIT) in 1995.  This photometry is
slightly deeper than that reported at 1620 \AA\ by Whitney et
al. (\cite{whro98}) from an earlier UIT flight. When selecting the
targets we concentrated on the stars at the very hot end of the blue
tail (cf. Fig.~\ref{ocen_kalu} and Table~1).  In order to observe as
many stars as possible we oriented the slit to cover two candidate
stars at once if possible.  This of course did not allow to reduce the
light loss due to atmospheric dispersion by observing along the
parallactic angle and also required observations in fairly crowded
regions.

We used grating \#4 (72~\AA$\,$mm$^{-1}$) with CCD \#31 (1024
$\times$ 1024 pixels of 24$\mu m^2$ size; 2.84 e$^-$/ADU,
read-out-noise 7.3 e$^-$) and a slit width of 1\bsec0, yielding a
spectral resolution of about 6.5 \AA\ as determined from the FWHM of
the wavelength calibration lines.  We observed each night ten bias
frames and ten dome flat fields and for the whole run two sky flat
fields to ensure a good correction of the illumination profile of the
slit.  At the beginning of each night we observed HeAr spectra for
wavelength calibration.  Due to the long integration times of the Ar
lamp we observed only He spectra before and after each science
exposure during the night, from which we derived the zero-points for
the wavelength calibration, while the dispersion relation was derived
from the HeAr frames.  We observed dark frames of 3600 and 1800 sec
duration to measure the dark current of the CCD. As flux standard star
we used Hiltner~600.

\begin{table*}
\caption{Positions, photometric information, and atmospheric parameters of 
target stars}
\begin{tabular}{l|ll|lll|lll}
\hline
\hline
Star & $\alpha_{2000}$ & $\delta_{2000}$ & $V$ & $V-I$ & $m_{1520}$
& \teff & \logg & \loghe \\
 & & & & & & [K]   & [cm s$^{-2}$] & \\
\hline
 WF3-1   &\RA{13}{26}{32}{5} & \DEC{$-$47}{24}{17}{7} & \magpt{18}{6} & & 
  & 35100$\pm$1600  & 5.80$\pm$0.28 & $-$2.18$\pm$0.59\\
BC6022 & \RA{13}{26}{24}{95} & \DEC{$-$47}{33}{23}{2} & \magpt{18}{494} & 
\magpt{$+$0}{057} & \magpt{15}{50}
 & 45600$\pm$1300 & 6.10$\pm$0.14 & $-$1.78$\pm$0.16\\
BC8117 & \RA{13}{26}{33}{23} & \DEC{$-$47}{35}{12}{3} & \magpt{18}{709} & 
\magpt{-0}{065} & \magpt{14}{81}
 & 29800$\pm$1000 & 5.48$\pm$0.14 & $-$2.30$\pm$0.23\\
BC21840 & \RA{13}{27}{25}{95} & \DEC{$-$47}{32}{19}{6} & \magpt{18}{103} & 
\magpt{-0}{061} & \magpt{15}{19}
 & 35700$\pm$\hspace*{1.1ex}700 & 5.55$\pm$0.14 & $-$0.80$\pm$0.14\\
C521 & \RA{13}{26}{08}{82} & \DEC{$-$47}{37}{12}{3} & \magpt{18}{676} & 
\magpt{-0}{078} & \magpt{15}{29}
 & 34700$\pm$\hspace*{1.1ex}500 & 5.90$\pm$0.12 & $-$0.90$\pm$0.09\\
 D4985 & \RA{13}{25}{14}{72} & \DEC{$-$47}{32}{36}{0} & \magpt{18}{867} & 
\magpt{-0}{078} & \magpt{15}{14}
 & 38400$\pm$\hspace*{1.1ex}800 & 6.08$\pm$0.16 & $-$0.87$\pm$0.16\\
D10123 & \RA{13}{25}{34}{25} & \DEC{$-$47}{29}{50}{0} & \magpt{18}{921} & 
\magpt{-0}{115} & \magpt{15}{21}
& 35000$\pm$\hspace*{1.1ex}500 & 5.82$\pm$0.12 & $-$0.87$\pm$0.09\\
D10763 & \RA{13}{25}{35}{57} & \DEC{$-$47}{27}{45}{1} & \magpt{19}{103} & 
\magpt{-0}{050} & \magpt{15}{46}
 & 35200$\pm$1500 & 4.35$\pm$0.19 & $+$0.94$\pm$0.14\\
D12564 & \RA{13}{25}{41}{34} & \DEC{$-$47}{29}{06}{0} & \magpt{19}{007} & 
\magpt{-0}{182} & \magpt{15}{13}
 & 36900$\pm$1000 & 5.60$\pm$0.14 & $-$0.37$\pm$0.09\\
D14695 & \RA{13}{25}{46}{54} & \DEC{$-$47}{26}{51}{7} & \magpt{19}{220} & 
\magpt{-0}{025} & \magpt{15}{59}
 & 41300$\pm$\hspace*{1.1ex}700 & 6.11$\pm$0.21 & $-$0.22$\pm$0.12\\
D15116 & \RA{13}{25}{50}{19} & \DEC{$-$47}{32}{05}{9} & \magpt{18}{760} & 
\magpt{-0}{080} & \magpt{14}{87}
 & 41500$\pm$1100 & 6.11$\pm$0.14 & $-$0.21$\pm$0.11\\
D16003 & \RA{13}{25}{53}{96} & \DEC{$-$47}{35}{21}{3} & \magpt{18}{891} & 
\magpt{-0}{067} & \magpt{15}{49}
 & 36300$\pm$\hspace*{1.1ex}600 & 5.91$\pm$0.12 & $-$1.03$\pm$0.10\\
\hline
\end{tabular}
\end{table*} 

The flat fields showed slight ($\le$ $\pm$2\%) variations on short
timescales ($\sim$ minutes), while the bias frames showed no
variations.  We therefore averaged the dome flat fields and the bias
frames for all nights.  The mean sky flat field and the mean dome flat
field were averaged along the dispersion axis to construct their
respective spatial illumination profiles.  The mean dome flat field
was then corrected by the ratio of the illumination profiles.  By
averaging the dome flat field along its spatial axis we determined the
spectral energy distribution of the flat field lamp and corrected it
by dividing the mean flat field through the heavily smoothed energy
distribution.

\begin{figure}[ht]
\vspace*{11.85cm}
\includegraphics{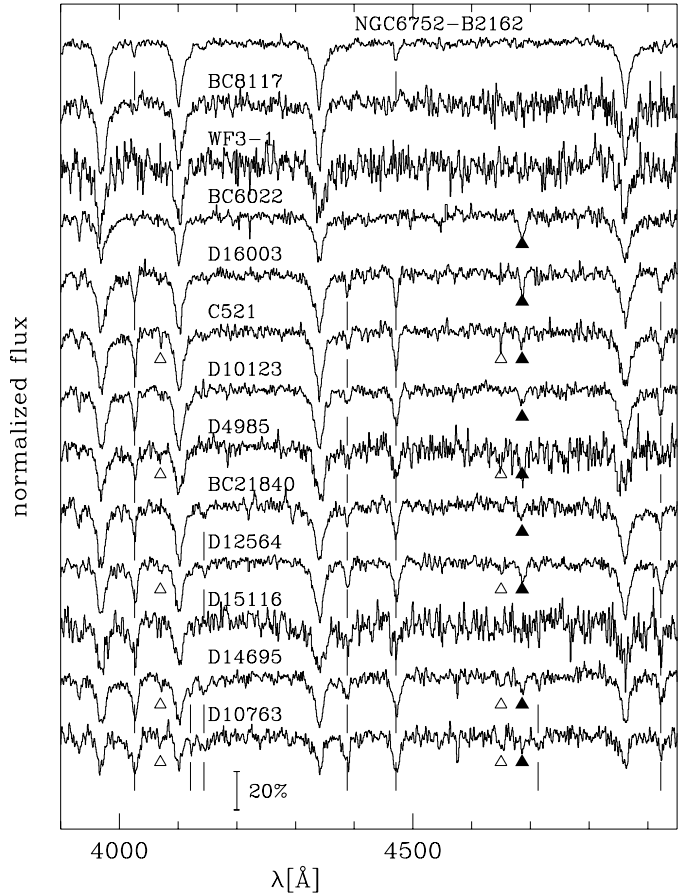}
\caption{Spectra of blue hook stars in $\omega$ Cen sorted
according to their helium abundance. For comparison
the spectrum of a star at the faint end of the blue tail in NGC~6752
from Moehler et al. (\cite{mosw00}) is shown. The lines mark the position of
the He~{\sc i} lines, the filled triangles mark the He~{\sc ii} line
at 4686\AA, the open triangles mark the C~{\sc iii} lines at 4070~\AA\
and the C~{\sc iii}/N~{\sc iii} lines at 4650\AA.
\label{ocen_spec}}
\end{figure}

For the wavelength calibration we fitted 2$^{\rm nd}$-order
polynomials to the dispersion relations of the HeAr spectra (using 17
unblended lines) which resulted in mean residuals of $\le$0.4~\AA. We
rebinned the frames two-dimensionally to constant wavelength
steps. The two-dimensional sky subtraction was performed as described
in Moehler et al. (\cite{mosw00}) with the spatial profile of the sky
background described by a constant.  The sky-subtracted spectra were
extracted using Horne's (\cite{horn86}) algorithm as implemented in
MIDAS (Munich Image Data Analysis System).
Finally the spectra were corrected for atmospheric extinction using
the extinction coefficients for La Silla (T\"ug \cite{tueg77}) as
implemented in MIDAS.  The flux data for Hiltner~600 were taken from
Hamuy et al. (\cite{hawa92}) and the response curves were fitted by
splines.  The flux-calibration is helpful for the later normalization
of the spectra as it takes out all large-scale sensitivity variations
of the instrumental setup. Atmospheric dispersion will cause light loss
especially at blue end of the spectral range so that the flux
distribution of the calibrated spectra cannot be used to infer
temperatures (e.g. from the Balmer jump).
We determined radial velocity shifts from the positions of the Balmer
and \ion{He}{i} lines.  The Doppler-corrected spectra were then
co-added and normalized by eye and are plotted in
Fig.~\ref{ocen_spec}.

\section{Analysis}

In contrast to the somewhat brighter (in absolute visual magnitude)
blue tail stars analysed in NGC~6752, which are helium deficient and
show weak to no helium lines (cf. Moehler et al. \cite{mosw00} and
uppermost spectrum in Fig.~\ref{ocen_spec}), most of the blue hook
stars in $\omega$~Cen show rather strong He~{\sc i} lines, and some of
them even show C~{\sc iii}/N{\sc iii} and He~{\sc ii} absorption (see
Fig.~\ref{ocen_spec}).

\begin{figure}[h]
\vspace*{11cm}
\includegraphics{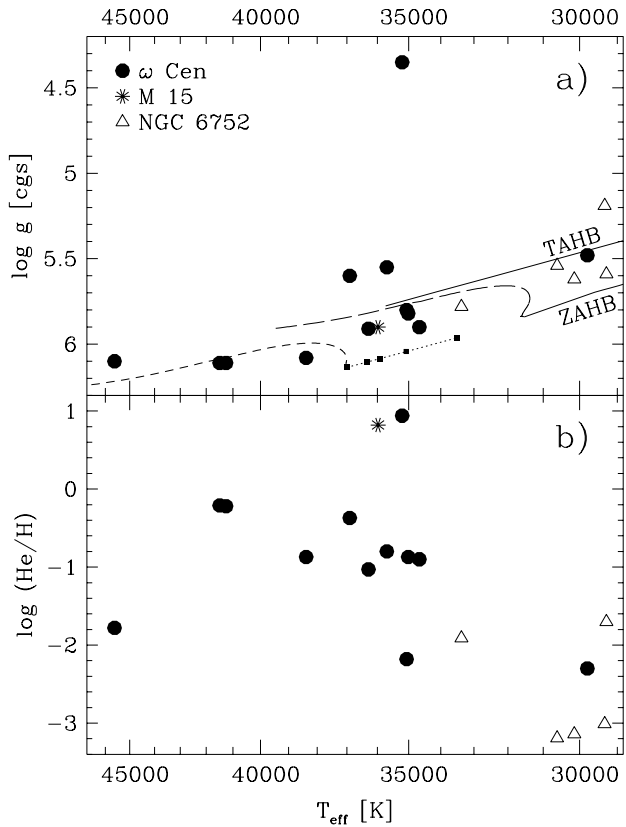}
\caption{{\bf a)} Atmospheric parameters 
derived from the spectra of blue hook
 stars in $\omega$ Cen compared to HB evolutionary tracks. Also
shown are blue tail stars from NGC~6752 (open triangles, Moehler et
al. \cite{mosw00}) and the helium-rich sdB star in M~15 (starry
symbol, Moehler et al. \cite{mohe97}).  The tracks for an early hot
flasher (long-dashed line) and a late hot flasher (short-dashed
line) show the evolution of such stars from the zero-age HB (ZAHB)
towards helium exhaustion in the core (terminal-age HB = TAHB). The
solid lines mark the canonical HB locus for [M/H] = $-$1.5 from
Sweigart (\cite{swei97}). The dotted line connects the series of ZAHB
models computed by adding a hydrogen-rich layer to the surface of the
ZAHB model of the late hot flasher.  The filled squares mark --
with decreasing temperature -- hydrogen layer masses of $0, 10^{-7},
10^{-6}, 10^{-5}, 10^{-4}$\Msolar.
{\bf b)} Helium abundances for the same stars. 
\label{ocen_tg}}
\end{figure}

Fits to the spectra with non-LTE model atmospheres allow to derive
effective temperatures, surface gravities, and helium abundances.  The
helium-rich NLTE model atmospheres were calculated with a
modified version of the accelerated lambda iteration code of Werner \&
Dreizler (\cite{wedr99}). The model atoms for hydrogen and helium as
well as the handling of the line broadening for the spectrum synthesis
are similar to those of Werner (\cite{wern96}).  The calculation of
the helium-poor NLTE model atmospheres is described in Napiwotzki
(\cite{napi97}). 
To establish the best fit we used the routines developed by
Bergeron et al.\ (\cite{besa92}) and Saffer et al.\ (\cite{sabe94}), as
modified by Napiwotzki et al. (\cite{nagr99}), 
which employ a $\chi^2$ test. The $\sigma$ necessary for the
calculation of $\chi^2$ is estimated from the noise in the continuum
regions of the spectra. The fit program normalizes model spectra {\em
and} observed spectra using the same points for the continuum
definition.  The results obtained from fitting the Balmer
lines H$_\beta$ to H$_{10}$ (excluding H$_\epsilon$ to avoid the
\ion{Ca}{ii}~H line), the \ion{He}{i} lines $\lambda\lambda$ 4026~\AA,
4388~\AA, 4471~\AA, 4921~\AA\, and the \ion{He}{ii} lines
$\lambda\lambda$ 4542~\AA, 4686~\AA\ are given in Table~1 and
plotted in Fig.~\ref{ocen_tg}.

Nine of the twelve stars show at least solar helium
abundance\footnote{\loghe$_\odot$ = $-$1} (as
opposed to the hottest EHB stars in NGC~6752, which show helium
abundances of $\le$0.1 solar, Moehler et al. \cite{mosw00}) and four have a
helium abundance by particle number of $\ge$0.4 (corresponding to $Y
\ge$0.7).  The only other globular cluster blue tail star which has
been found to show a super-solar helium abundance is M15 F2-2 (Moehler
et al. \cite{mohe97}), which is also quite hot ($T_{\rm eff}\approx$
36,000 K). 

Synthetic NLTE spectra suggest that a somewhat super-solar carbon
abundance of [C/H] $\approx +0.5\pm0.5$ is required
to explain the CIII features, although a quantitative analysis will
require higher quality spectra. 
As noted in Sect.~2, an enhanced
carbon abundance in the blue hook stars is predicted by the
flash-mixing scenario.

\section{Discussion}
Our analysis of the blue hook stars in $\omega$~Cen shows that these
stars do indeed reach effective temperatures of more than 35,000 K
(cf. Fig.~\ref{ocen_tg} and Table~1),
well beyond the hot end of the canonical EHB.  In addition, most of
them show at least solar helium abundances with the helium abundance
increasing with effective temperature (cf. Fig.~\ref{ocen_tg}), in
contrast to canonical EHB stars such as those studied in NGC~6752 by
Moehler et al. (\cite{mosw00}).  We now discuss both of these results
in more detail.

The coolest star in our sample (BC8117) at $T_{\rm eff} \approx 30,000$ K
lies near the hot end of the canonical EHB and shows the same low
helium abundance as the EHB stars in NGC~6752 (see Fig.~\ref{ocen_tg}).  Most
likely, BC8117 is the descendant of an early hot flasher.  All of
the other stars in our sample have temperatures $\gtrsim$35,000 K
and, except for the low gravity star D10763, lie in the general
vicinity of the track for a late hot flasher in Fig.~\ref{ocen_tg}.
Although limited, our data suggest that the
blue hook stars may be separated from the canonical EHB
stars by a temperature gap from $\approx$31,000 K to $\approx$35,000 K.  As
discussed in Section 2, such a temperature gap is predicted by
the flash-mixing scenario.

\begin{figure}[h]
\vspace*{8cm}
\includegraphics{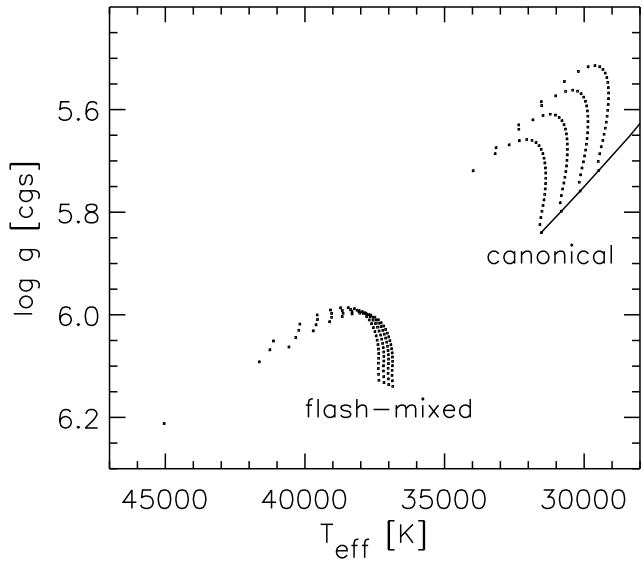}
\caption{HB evolutionary tracks for 4 canonical stars near the
hot end of the EHB and 4 late hot flashers from
Brown et al. (\cite{brsw01}).  Each track is represented by
a series of points separated by a time interval
of $5 \times 10^6$ yr.  The solid line is the canonical
ZAHB.  Note the temperature gap between the canonical and
flash-mixed tracks.
\label{track_time}}
\end{figure}

The HB track for the early hot flasher in Fig.~\ref{ocen_tg}
passes through the temperature gap, thus raising the possibility that
canonical EHB stars might populate this gap during their post-ZAHB
evolution. To examine this possibility more closely, we plot in
Fig.~\ref{track_time} the HB evolutionary tracks from Brown et
al. (\cite{brsw01}) for 4 canonical stars near the hot end of the EHB
and 4 late hot flashers.  The latter tracks span the range in RGB
mass loss over which flash mixing occurs.  Each track is represented
by a series of points separated by a time interval of $5 \times 10^6$
yr in order to illustrate where the evolution is slowest.
Fig.~\ref{track_time} shows that canonical EHB stars spend almost
their entire HB lifetime at temperatures close to their ZAHB
temperatures.  While these stars evolve into the temperature gap near
the end of the HB phase, they do so at a time when their evolution is
very rapid.  Thus one would not expect to find many evolved EHB stars
within the temperature gap or along the part of the terminal-age HB
(TAHB) that extends into the temperature gap in Fig.~\ref{ocen_tg}.
We conclude that the flash-mixed stars should remain well separated in
temperature from the canonical EHB stars also when HB evolution is taken
into account.

Contrary to our original expectations, the atmospheres of the
blue hook stars still show some hydrogen.  This result may be
understood in light of the recent calculations of Schlattl \& Weiss
(2002, priv. comm.), who found that a small amount of hydrogen
survives the flash mixing.  The observed atmospheric hydrogen
abundance of the blue hook stars is, however, substantially greater
than the predicted envelope hydrogen abundance (X $\approx$ 10$^{-4}$)
in the models of Schlattl \& Weiss after flash mixing.  This apparent
discrepancy could be readily explained by the outward diffusion of
hydrogen into the atmospheres of the blue hook stars and the
gravitational settling of helium.  Such diffusive processes are
believed to be responsible for the low helium abundances of the sdB
stars and are estimated to operate on a time scale much shorter than
the HB lifetime.  The range in the hydrogen abundances of the
blue hook stars might indicate that varying amounts of hydrogen
survive flash mixing or that the efficiency of diffusion differs
from star to star.  In any case the high helium abundances observed in
some of the blue hook stars would be difficult to understand if their
atmospheres were not enriched in helium during the helium flash.  The
increase in the mean atmospheric helium abundance with increasing
effective temperature is also consistent with flash mixing.

The presence of a hydrogen-rich surface layer would shift the
evolutionary track for the late hot flasher in Fig.~\ref{ocen_tg}
towards cooler temperatures.  This evolutionary track, taken from the
blue hook sequences of Brown et al. (\cite{brsw01}), has a
helium/carbon-rich envelope with no hydrogen.  In order to estimate
the size of this temperature shift, we computed a series of ZAHB
models in which hydrogen-rich layers with masses of 10$^{-7}$,
10$^{-6}$, 10$^{-5}$ and 10$^{-4}$~M$_\odot$ were added to the ZAHB
model from the late hot flasher in Fig.~\ref{ocen_tg}.  A hydrogen
layer of 10$^{-4}$~M$_\odot$ corresponds to the case in which
$\approx$10 percent of the envelope hydrogen survives flash mixing and
in which all of this hydrogen then diffuses to the surface.  This
should be a firm upper limit to the mass of any hydrogen layer, given
the results of Schlattl \& Weiss (2002, priv. comm.) and the fact that
any hydrogen present in the deeper layers of the envelope would not
have sufficient time to diffuse to the surface during the HB phase.
As expected, the ZAHB location of the late hot flasher in
Fig.~\ref{ocen_tg} shifts redward as the mass of the hydrogen layer
increases and we see that the addition of a hydrogen layer of $<$
10$^{-4}$~M$_\odot$ would actually improve the agreement between the
predicted and observed temperatures of the blue hook stars while at
the same time preserving the temperature gap between these stars and
the canonical EHB stars.

A most intriguing puzzle is posed by D10763, which is the most
helium-rich star in our sample: While it is among the faintest stars
visually, its low surface gravity suggests a very high luminosity,
which would put it to a distance of about 50~kpc for a mass of
0.5\Msolar. Its heliocentric radial velocity of $+170\pm40$~km
s$^{-1}$, however, suggests that it is a member of $\omega$ Cen.
The spectrum also shows no evidence for features from a cool star
(e.g., stronger \ion{Ca}{ii} K line or G band), which might influence
the parameter determination from the Balmer lines. We have currently
no explanation for this object.

\section{Conclusions}
The high temperatures and high helium and carbon abundances
reported here for the blue hook stars in $\omega$ Cen provide general
support for the flash-mixing hypothesis of Brown et
al. (\cite{brsw01}).  However, several questions remain. 

The CMD of $\omega$ Cen of Kaluzny et al.\ (\cite{kaku97}) given in 
Fig.~\ref{ocen_kalu}
does not show clear evidence for a gap within the EHB such as
was found in NGC 2808 at M$_V \approx$ \magpt{+4}{6} by Walker 
(\cite{walk99}) and
Bedin et al. (\cite{bepi00}).  Brown et al. (\cite{brsw01}) have shown that
the EHB gap in NGC 2808 can be identified with the
transition between the canonical and flash-mixed stars.  There
is a gap at M$_V \approx$ \magn{+4} in Fig.~\ref{ocen_kalu}, 
but this gap separates the
canonical EHB from blue HB stars and is not related to the hot
flasher scenario (D'Cruz et al. \cite{dcoc00}).  A fuller discussion of the
gap between the EHB and blue HB in a number of other globular clusters
is given by Piotto et al. (\cite{pizo99}). One possible reason for the
absence of a clear EHB gap in the $\omega$ Cen CMD may be the
limited precision of the photometry of Kaluzny et
al. (\cite{kaku97}), who warn about possible problems at faint
magnitudes.  Alternatively, the absence of a clear gap may be related
to the metallicity spread  in $\omega$ Cen, although the models of D'Cruz et
al. (\cite{dcdo96}; their Fig.~2) suggest that the temperature at the hot end
of the EHB shows little dependence on metallicity.  Given the known radial
metallicity gradient in $\omega$ Cen (e.g.\ Hilker \&
Richtler \cite{hiri00}), it would be of interest to determine if there
is a gradient in the fraction of EHB stars which are blue hook stars.  

Another
question is why flash-mixed stars appear in $\omega$ Cen and NGC~2808
but not in other EHB clusters such as M~13 and NGC~6752.  
Both $\omega$ Cen (M$_V$ = \magpt{-10}{29}; Harris \cite{harr96}) and
NGC~2808 (M$_V$ = \magpt{-9}{36}) are among the most massive
globular clusters in the Galaxy, so that the observed large EHB population in
these clusters is not unexpected. However, the question of why a larger
fraction of EHB stars in these clusters should be blue hook stars remains
unexplained, and can be considered as another twist in the general problem of
understanding the origin of the HB morphologies in globular clusters.

\begin{acknowledgements} 
We want to thank the staff at La Silla for their support during our
observations and R. Napiwotzki for his model atmospheres. 
SM was supported by the DLR under grant
50\,OR\,96029-ZA. AVS gratefully acknowledges 
support from NASA Astrophysics Theory Proposal
NRA-99-01-ATP-039.  SD thanks Nicolay Hammer,
Agnes Hoffmann, Iris Traulsen for their help
calculating the model atmosphere grid.
\end{acknowledgements}

\end{document}